\documentclass [aps, prb, twocolumn,floatfix,10pt]{revtex4}%
\usepackage{graphicx}
\usepackage{bm,amsmath,amssymb,mathrsfs,dcolumn}
\usepackage [colorlinks=true,linkcolor=blue,citecolor=blue,urlcolor=blue,linktoc=page,bookmarks=false,pdfstartview={FitH},pdfborder={0 0 0.0 [3 3]}]{hyperref}
\usepackage [usenames]{xcolor}
\hypersetup{pdfborder=0 0 0,colorlinks=true,citecolor=blue,linkcolor=blue}
\bibpunct{[}{]}{,}{n}{}{}

\def\be{\begin{equation}}
\def\ee{\end{equation}}
\def\ber{\begin{eqnarray}}
\def\eer{\end{eqnarray}}
\def\bs{\boldsymbol}
\usepackage{setspace}
\begin{document}
\title{Magnon spin transport around the compensation magnetic field in easy-plane antiferromagnetic insulators}
\author {Ka Shen}
\thanks{kashen@bnu.edu.cn}
\affiliation{The Center for Advanced Quantum Studies and Department of Physics, Beijing Normal University, Beijing 100875, China}
\date{\today }


\begin{abstract}
  In this work, we theoretically study the magnon spin transport in easy-plane antiferromagnetic insulators in the presence of an in-plane magnetic field. By exactly calculating the magnon spectrum, we find the band splitting due to the magnetic anisotropy can be fully compensated by the external field at a particular strength, which makes its dynamics nearly equivalent to an easy-axis antiferromagnet. As a result, the intrinsic magnon spin Hall effect due to the dipole-dipole interaction, previously predicted in easy-axis antiferromagnets is activated in easy-plane antiferromagnets. The compensation feature also allows the field control of magnon spin lifetime and hence the spin diffusion lenth. The compensation feature is robust against the biaxial anisotropy.


\end{abstract}
\maketitle

\section{Introduction}
The magnetic dynamics and transport properties in antiferromagnetic materials are essential for the performance of antiferromagnetic spintronic devices, and have been investigated intensively in the past decade~\cite{Baltz18,Hoffmann18}. Among these studies, high efficient spin transmission through an insulating antiferromagnetic layer was demonstrated in ferromagnet-antiferromagnet-normal metal (NM) trilayer structures~\cite{WangHL2014,Lin16,Qiu16,YiWang2019}. The lack of itinerant electrons implies that the spin information should be able to transmit across the antiferromagnetic insulator (AFI) in form of polarized magnons. Non-local measurement in different materials revealed that the spin transport distance associated with antiferromagnetic magnons can reach several or even tens microns~\cite{Lebrun18,Yuan2018,2D_Xing19}, which is already comparable with that in high quality yttrium iron garnet, a ferrimagnetic material famous for its ultralow magnetic damping~\cite{Cornelissen15,Cornelissen16b}. Recently, spin injection into NM from subterahertz magnons  was realized  in AFI-NM bilayers by spin pumping~\cite{JShi20,Vaidya20} and optical approach~\cite{DiWu20}. These progresses offer new opportunities for promising applications of AFIs.

As most of the previous experimental works are about easy-axis AFIs, such as $\alpha$-Fe$_2$O$_3$ (below transition temperature around 260~K in bulk)~\cite{Lebrun18}, Cr$_2$O$_3$~\cite{Yuan2018,Qiu18,JShi20}, MnF$_2$~\cite{Vaidya20}, and MnPS$_3$~\cite{2D_Xing19}, systems with a magnetic easy plane like NiO and $\alpha$-Fe$_2$O$_3$ (above transition temperature) are also quite interesting because of their distinctive features. For instance, the U(1) spin-rotational symmetry within the easy plane allows spin superfluidity~\cite{Halperin69,Takei14,Qaium17}. Another important advantage for applications, compared with the easy-axis AFIs, is the easy access of magnetization manipulation, because the Neel vector in easy-plane AFIs keeps perpendicular to and therefore can be rotated by an in-plane magnetic field. This allows field modulation of spin transport~\cite{Luqiao20} and spin Hall magnetoresistance~\cite{Wees20}. In contrast to the easy-axis case, where the magnon bands are two-fold degenerate, the magnetic anisotropy in easy-plane AFIs breaks the symmetry between the in-plane and out-of-plane magnetization dynamics and results in a band splitting. The lower and higher frequency modes corresponds to the in-plane and out-of-plane motion of the Neel vector~\cite{Rezende19}, respectively. Such a splitting leads to a coherent dynamics of magnon spin polarization~\cite{Luqiao20} and its modulation by an external magnetic field causes a Hanle-type effect~\cite{Wimmer20,Kamra20}.

Interestingly, the spatial motion of the magnons in AFIs, similar to the mobile electrons in metallic systems, can be correlated with their spin polarization, for example, in noncentrosymmetric systems via the Dzyaloshinskii–Moriya interaction (DMI), which provides the possibility to discovery electron-like spin-orbit phenomena. Theoretical studies predicted magnon spin Nernst effect~\cite{Cheng2016,Kovalev16} and magnonic Edelstein effect~\cite{Kovalev20,Rancheng20} driven by DMI. Recently, the dipole-dipole interaction (DDI), which was usually ignored in antiferromagnets, was shown to be able to manifest itself as an effective spin-orbit coupling (SOC)~\cite{Shen20,JLiu20b} between magnon states  in uniaxial easy-axis AFIs. Such a magnon SOC can also give rise to various spin-orbit phenomena, e.g., an intrinsic magnon spin Hall effect (SHE)~\cite{Shen20}, D'yakonov-Perel' (DP)-type magnon spin relaxation, and topological surface states~\cite{JLiu20}. The role of this DDI-induced mechanism in easy-plane AFIs is yet to be examined.

In this work, we calculate the magnon spectrum analytically in easy-plane AFIs by taking into account the exchange interaction, magnetic anisotropy, Zeeman energy due to an in-plane magnetic field and, as interpreted above, the DDI. Since the magnitude of DDI is relatively weak than the splitting between the in-plane and out-of-plane polarized magnon modes under magnetic anisotropy, its influence is negligible in weak magnetic field regime. Very interestingly, as the magnetic field increases, the band splitting is found to be globally suppressed and, at a compensation magnetic field, the contributions from the magnetic anisotropy and Zeeman term cancel with each other exactly in the entire Brillouin zone, making the easy-plane AFI approximately equivalent to an easy-axis one.
Physically, this is because the external field introduces an additional magnetic anisotropy, which behaves as a hard axis along the magnetic field and together with the original natural hard axis defines a hard plane, making the normal direction equivalently an easy axis. The resulting magnetic anisotropy becomes uniaxial when the strengths in the two hard axes are equal.
As a result, the momentum-dependent SOC due to DDI becomes dominant and the magnon spin Hall mechanism is switched on. In the meantime, the DP-type magnon spin relaxation, although it is relevant regardless of the strength of magnetic field, can be strongly modified around the compensation field. Moreover, the role of DMI and the additional magnetic anisotropy within the easy plane will also be addressed. 


\section{Model}\label{smodel}
\begin{figure}[tp]
  \includegraphics[width=5cm]{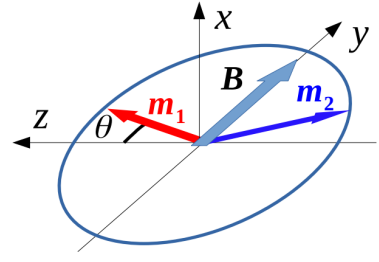}
  \caption{The spin configuration of an easy-plane AFI at equilibrium state in the presence of an in-plane external magnetic field.}
  \label{config}
\end{figure}

We start from the minimal model for an easy-plane AFI including the magnetic anisotropy and antiferromagnetic exchange interaction between the nearest neighbors. An external magnetic field is applied within the $y$-$z$ easy plane to control the orientation of the Neel vector. Without loss of generality, as illustrated in Fig.~\ref{config}, the magnetic field is set to be along $y$-axis, which leads to a canting of the two antiferromagnetic coupled sublattice magnetizations $\bs m_1$ and $\bs m_2$. The net magnetization $\bs m=\bs m_1 +\bs m_2$ and  the Neel vector $\bs n=\bs m_1 -\bs m_2$ are therefore along $y$ and $z$ directions, respectively. The canting angle $(\theta)$ can be determined by minimizing the total energy described by the Hamiltonian
\ber
H&=&\sum_{i}\left[K(S_{ai}^{x})^{2}+K(S_{di}^{x})^{2}-g\mu_{B}BS_{ai}^{y}-g\mu_{B}BS_{di}^{y}\right]\nonumber\\
&&-\sum_{\langle i,j\rangle}J\boldsymbol{S}_{ai}\cdot\boldsymbol{S}_{dj},
\label{H0}
\eer
where the anisotropy coefficient $K>0$ and the inter-sublattice exchange coupling constant $J<0$.
The subscripts $a$ and $d$ label the two sublattices.
For a system with $2N$ magnetic ions, the total energy reads
\be
E\approx-Nz|J|S^{2}\cos2\theta-2Ng\mu_{B}BS\sin\theta,
\label{totalE}
\ee
and thus the canting angle is determined by
\be
\sin\theta=\frac{\omega_{\rm Z}}{2\omega_{\rm ex}}.
\ee
Here, $\omega_{\rm Z}=g\mu_B B/\hbar$ and $\omega_{\rm ex}=z|J|S/\hbar$ represent the frequency scales of the Zeeman term and the exchange interaction, respectively.

The spin operators in Eq.~(\ref{H0}) can be expressed under the local equilibrium configuration via a rotation operation
\ber
\left(\begin{array}{c}
  S_{ai}^{x}\\
  S_{ai}^{y}\\
S_{ai}^{z}
\end{array}\right)	&=&	\left(\begin{array}{ccc}
  1 &0 &0\\
0& \cos\theta & \sin\theta\\
0& -\sin\theta & \cos\theta
\end{array}\right)\left(\begin{array}{c}
  \tilde{S}_{ai}^{x}\\
  \tilde{S}_{ai}^{y}\\
\tilde{S}_{ai}^{z}
\end{array}\right),\\
\left(\begin{array}{c}
  S_{di}^{x}\\
  S_{di}^{y}\\
S_{di}^{z}
\end{array}\right)	&=&	\left(\begin{array}{ccc}
  1 &0 &0\\
0& \cos\theta & -\sin\theta\\
0& \sin\theta & \cos\theta
\end{array}\right)\left(\begin{array}{c}
  \tilde{S}_{di}^{x}\\
  \tilde{S}_{di}^{y}\\
S_{di}^{z}
\end{array}\right),
\eer
which leads to
\ber
H&=&\sum_{i}\left[K(\tilde{S}_{ai}^{x})^{2}+K(\tilde{S}_{di}^{x})^{2}-g\mu_{B}B\sin\theta(\tilde{S}_{ai}^{z}-\tilde{S}_{di}^{z})\right]\nonumber\\
&&-\sum_{\langle i,j\rangle}J\left[\tilde{\boldsymbol{S}}_{ai}\cdot\tilde{\boldsymbol{S}}_{dj}-2\sin^{2}\theta(\tilde{S}_{ai}^{y}\tilde{S}_{dj}^{y}+\tilde{S}_{ai}^{z}\tilde{S}_{dj}^{z})\right.\nonumber\\
  &&\hspace{1cm}\left. -\sin2\theta(\tilde{S}_{ai}^{y}\tilde{S}_{di}^{z}-\tilde{S}_{ai}^{z}\tilde{S}_{di}^{y})\right].\label{EqH0}
\eer
By performing the standard Holstein-Primakoff transformation~\cite{Holstein40} to the spin operators
\ber
\tilde S_{a}^{z}=S-a^{\dagger}a,& \tilde S_{a}^{+}=\sqrt{2S-a^{\dagger}a}a,\nonumber\\
\tilde S_{d}^{z}=-S+d^{\dagger}d,& \tilde S_{d}^{+}=d^{\dagger}\sqrt{2S-d^{\dagger}d},
\label{HP}
\eer
one can write out the quadratic terms  in momentum space under the basis of $(a_{\boldsymbol{k}},d_{\boldsymbol{k}},a_{-\boldsymbol{k}}^{\dagger},d_{-\boldsymbol{k}}^{\dagger})^T$ as
\be
H_{\bs k,-\bs k}^0=
\left(\begin{array}{cccc}
{\cal A} & {\cal C}_{\boldsymbol{k}} & {\cal {\cal A}}' & {\cal B}_{\boldsymbol{k}}-{\cal C}_{\boldsymbol{k}}\\
{\cal C}_{\boldsymbol{k}} & {\cal A} & {\cal B}_{\boldsymbol{k}}-{\cal C}_{\boldsymbol{k}} & {\cal {\cal A}}'\\
{\cal {\cal A}}' & {\cal B}_{\boldsymbol{k}}-{\cal C}_{\boldsymbol{k}} & {\cal A} & {\cal C}_{\boldsymbol{k}}\\
{\cal B}_{\boldsymbol{k}}-{\cal C}_{\boldsymbol{k}} & {\cal {\cal A}}' & {\cal C}_{\boldsymbol{k}} & {\cal A}
\end{array}\right),\label{H00}
\ee
where ${\cal A}/\hbar=\omega_{{\rm an}}+\omega_{{\rm ex}}$, ${\cal A}'/\hbar=\omega_{{\rm an}}$, $\cal B_{\bs k}/\hbar=\gamma_{\boldsymbol{k}}\omega_{{\rm ex}}$, and ${\cal C}_{\boldsymbol{k}}/\hbar=\gamma_{\boldsymbol{k}}\omega_{\delta}$ with $\omega_{\rm an}={KS}/\hbar$ and $\omega_{\delta}=\omega_{\rm ex}\sin^2\theta$. The form factor $\gamma_{\boldsymbol k}=(1/z)\sum_{\bs \delta}\exp(i \bs \delta \cdot \bs k)$ averages the phase factor over all $z$ antiferromagnetic coupled neighbors and is real in cubic lattice.

\subsection{Magnon dispersion relation}
In order to compute the dispersion relation analytically, it is convenient to define the magnon operators,  according to the symmetry, as orthogonal linearly polarized basis
\be
\phi^\pm_{\boldsymbol{k}}	=	(a_{\boldsymbol{k}}\pm d_{\boldsymbol{k}})/\sqrt{2},
\ee
and rewrite Hamiltonian (\ref{H00}) under the basis of $[\phi^+_{\boldsymbol{k}},(\phi^+_{-\boldsymbol{k}})^{\dagger},\phi_{\boldsymbol{k}}^-,(\phi^-_{-\boldsymbol{k}})^{\dagger}]^T$ as
\be
\tilde H_{\bs k,-\bs k}^0=
\left(\begin{array}{cccc}
{\cal A}+{\cal C}_{\boldsymbol{k}} & {\cal B}_{\boldsymbol{k}}^{+}-{\cal C}_{\boldsymbol{k}} & 0 & 0\\
{\cal B}_{\boldsymbol{k}}^{+}-{\cal C}_{\boldsymbol{k}} & {\cal A}+{\cal C}_{\boldsymbol{k}} & 0 & 0\\
0 & 0 & {\cal A}-{\cal C}_{\boldsymbol{k}} & {\cal B}_{\boldsymbol{k}}^{-}+{\cal C}_{\boldsymbol{k}}\\
0 & 0 & {\cal B}_{\boldsymbol{k}}^{-}+{\cal C}_{\boldsymbol{k}} & {\cal A}-{\cal C}_{\boldsymbol{k}}
\end{array}\right),
\label{Hkk}
\ee
in which ${\cal B}_{\boldsymbol{k}}^{\pm}=\cal A' \pm\cal B_{\bs k}$. Apparently, Hamiltonian~(\ref{Hkk}) can be divided into two individual BdG blocks, both of which can be solved analytically via the Bogoliubov transformation. Straightforward calculation gives the eigenfrequencies of two linear polarized magnon modes
\be
\omega_{\boldsymbol{k}}^{\pm}=\sqrt{({\cal A}+{\cal B}_{\boldsymbol{k}}^{\pm})({\cal A}-{\cal B}_{\boldsymbol{k}}^{\pm}\pm2{\cal C}_{\boldsymbol{k}})}/\hbar.
\label{freqpm}
\ee
and the operators of the eigenstates
\be \psi_{\boldsymbol{k}}^{\pm}	=	u_{\boldsymbol{k}}^{\pm}\phi_{\boldsymbol{k}}^{\pm}+v_{\boldsymbol{k}}^{\pm}(\phi_{-\boldsymbol{k}}^{\pm})^{\dagger},\label{psipm}
\ee
where the coefficients can be expressed by
\ber
u_{\boldsymbol{k}}^{\pm}	&=&	\sqrt{\frac{{\cal A}\pm{\cal C}_{\boldsymbol{k}}+\hbar\omega_{\boldsymbol{k}}^{\pm}}{2\hbar\omega_{\boldsymbol{k}}^{\pm}}},\\
v_{\boldsymbol{k}}^{\pm}	&=&	{\rm sgn}{({\cal B}_{\boldsymbol{k}}^{\pm}\mp{\cal C}_{\boldsymbol{k}})}\sqrt{\frac{{\cal A}\pm{\cal C}_{\boldsymbol{k}}-\hbar\omega_{\boldsymbol{k}}^{\pm}}{2\hbar\omega_{\boldsymbol{k}}^{\pm}}}.
\eer

In the long-wavelength limit, $\bs k\simeq 0$, one has $\gamma_{\bs k}\simeq 1$ and therefore
\ber
\omega_{\boldsymbol{0}}^{+}	&=&	2\sqrt{\omega_{\delta}(\omega_{{\rm ex}}+\omega_{{\rm an}})}=\omega_{Z}\sqrt{\frac{\omega_{{\rm ex}}+\omega_{{\rm an}}}{\omega_{{\rm ex}}}},\\
\omega_{\boldsymbol{0}}^{-}	&=&	2\sqrt{\omega_{{\rm an}}(\omega_{{\rm ex}}-\omega_{\delta})}=\sqrt{\frac{\omega_{{\rm an}}}{\omega_{{\rm ex}}}(4\omega_{{\rm ex}}^{2}-\omega_{Z}^{2})}.
\eer
Notice that $\omega_{\boldsymbol{0}}^{+}$ is proportional to the external field and therefore vanishes at zero field, whereas $\omega_{\boldsymbol{0}}^{-}$ is of finite value and relatively insensitive to the magnetic field. As a consequence, they become equal at
\be
\omega_{\delta }=\frac{\omega_{{\rm ex}}\omega_{{\rm an}}}{\omega_{{\rm ex}}+2\omega_{{\rm an}}},
\ee
corresponding to the compensation Zeeman field
\be
\omega_{Zc}=2\omega_{{\rm ex}}\sqrt{\frac{\omega_{{\rm an}}}{\omega_{{\rm ex}}+2\omega_{{\rm an}}}}.
\ee
In hematite, the compensation field is around $8$~T~\cite{Wimmer20}.
The canting angle at this compensation field is
\be
\sin\theta_c=\frac{\omega_{Zc}}{2\omega_{{\rm ex}}}=\sqrt{\frac{\omega_{{\rm an}}/\omega_{\rm ex}}{1+2\omega_{{\rm an}}/\omega_{\rm ex}}}.
\ee
Since the in-plane anisotropy in typical antiferromagnets is much smaller than the exchange energy, i.e., $\omega_{{\rm an}}/\omega_{\rm ex}\ll 1$, the canting angle $\theta$ at the compensation field is relatively small, retaining collinear antiferromagnetic configuration approximately.

\begin{figure}[tp]
  \includegraphics[width=8.5cm]{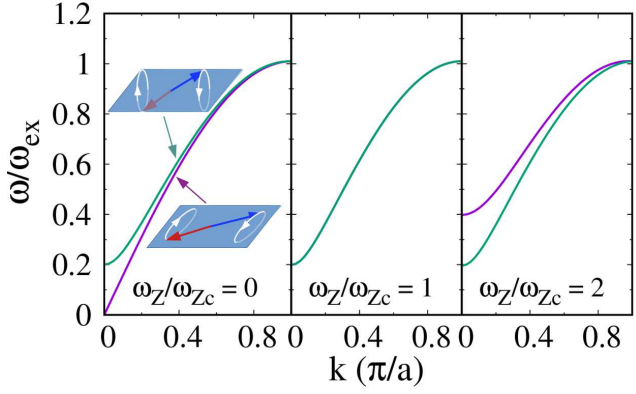}
  \caption{Dispersion relations of the two magnon modes (the green curve for $\omega^-_{\bs k}$ and the purple one for $\omega^+_{\bs k}$) with three typical magnetic field strengths. In the calculation, we adopt $\omega_{\rm an}/\omega_{\rm ex}=0.01$ and the form factor $\gamma_{\boldsymbol k}=\cos(k_xa/2)\cos(k_ya/2)\cos(k_ya/2)$ is applied along $(k_x,0,0)$ momentum line. The insets illustrate the magnetization dynamics of the two sublattices, with the white curves representing the trajectory of each magnetic moment.}
  \label{dispersion}
\end{figure}

Figure~\ref{dispersion} shows the dispersion relations with three typical strengths of the external magnetic field. The gapless linear dispersive mode ($\omega_{\bs k}^{+}$) branch in the absence of magnetic field corresponds to the Neel vector oscillating within the easy plane together with a small net magnetization oscillating out of the plane. In contrast, the gaped mode ($\omega_{\bs k}^-$) displays a large out-of-plane oscillation of the Neel vector along with a small in-plane magnetization oscillation. Since the $\omega_{\bs k}^{+}$ mode is more sensitive to the magnetic field than the $\omega_{\bs k}^{-}$ mode as discussed above, the two frequencies at $\bs k=0$ becomes equal at $\omega_Z=\omega_{Zc}$. Very importantly, according to the middle plot of Fig.~\ref{dispersion} and Eq.~(\ref{freqpm}), the two branches at this condition actually become degenerate for any $\bs k$. The dispersion relation reads
\be
\omega_{\boldsymbol{k}}^{\pm}=\hbar\sqrt{\frac{\omega_{{\rm ex}}}{\omega_{{\rm ex}}+2\omega_{{\rm an}}}}\sqrt{(2\omega_{{\rm an}}+\omega_{{\rm ex}})^{2}-(\gamma_{\boldsymbol{k}}\omega_{{\rm ex}})^{2}}.
\ee
Under this condition, an arbitrary combination of the two linearly polarized mode remains the eigenmode of Hamiltonian~(\ref{Hkk}), which allows a transform from the linear polarized modes to circularly polarized modes. This is very similar to the situation in easy-axis AFIs. In other words, the compensation magnetic field drives the easy-plane AFI into a configuration equivalent to an easy-axis AFI. This issue will be discussed further below in Sec.~\ref{Rspin}.
And, this effect is robust against an in-plane anisotropy as will be shown later in the paper.
For a field stronger than $\omega_{Zc}$, the $\omega_{\bs k}^{+}$ branch is lifted above the $\omega_{\bs k}^{-}$ one.

\subsection{DDI-induced SOC}
The long-range dipole-dipole interaction includes the coupling between any spin pair and it reads
\be
H^d=\frac{\mu_0 (g\mu_B)^2}{2}\sum_{l\ne l'} \frac{{|{\boldsymbol R}_{ll'}|}^2 \boldsymbol S_l\cdot\boldsymbol S_{l'}-3({{\boldsymbol R}_{ll'}}\cdot \boldsymbol S_l)({{\boldsymbol R}_{ll'}}\cdot \boldsymbol S_{l'})}{|{\boldsymbol R}_{ll'}|^5},\label{dd}
\ee
where $g$ is the $g$ factor, $\mu_B$ the Bohr magneton, and $\mu_0$ the vacuum permeability. As mentioned above, the ground spin configuration remains approximately collinear in the regime we are interested, due to the small canting angle. Therefore,
for the sake of simplicity, we make an approximation ${\bs S} \simeq \tilde {\bs S} $ to Eq.~(\ref{dd}) and apply the Holstein-Primakoff transformation, which results in,
under the basis of $(a_{\boldsymbol{k}},d_{\boldsymbol{k}},a_{-\boldsymbol{k}}^{\dagger},d_{-\boldsymbol{k}}^{\dagger})^T$~\cite{Shen20}
\be
H_{\boldsymbol k,-\bs k}^d=
\left(\begin{array}{cccc}
A_{\boldsymbol{k}} & \gamma_{\boldsymbol{k}}B_{\boldsymbol{k}}^{\ast} & B_{\boldsymbol{k}}^{\ast} & \gamma_{\boldsymbol{k}}A_{\boldsymbol{k}}\\
\gamma_{\boldsymbol{k}}B_{\boldsymbol{k}} & A_{\boldsymbol{k}} & \gamma_{\boldsymbol{k}}A_{\boldsymbol{k}} & B_{\boldsymbol{k}}\\
B_{\boldsymbol{k}} & \gamma_{\boldsymbol{k}}A_{\boldsymbol{k}} & A_{\boldsymbol{k}} & \gamma_{\boldsymbol{k}}B_{\boldsymbol{k}}\\
\gamma_{\boldsymbol{k}}A_{\boldsymbol{k}} & B_{\boldsymbol{k}}^{\ast} & \gamma_{\boldsymbol{k}}B_{\boldsymbol{k}}^{\ast} & A_{\boldsymbol{k}}
\end{array}\right),
\label{Hd}
\ee
in which
  \ber
A_{\boldsymbol{k}}&=&-2S\mu_{0}\mu_{B}^{2}\sum_{R_{ll'}\ne0} \frac{R_{ll'}^{2}-3(R_{ll'}^{z})^{2}}{R_{ll'}^{5}} e^{-i\mathbf{k}\cdot R_{ll'}},\\
B_{\boldsymbol{k}}&=&-6S\mu_{0}\mu_{B}^{2}\sum_{R_{ll'}\ne0} \frac{1}{R_{ll'}^{5}}(R_{ll'}^{+})^{2} e^{i\mathbf{k}\cdot R_{ll'}}.
\eer
  After computing the summation in continuum limit~\cite{Akhiezer68,Akashdeep17B,Shen20}, we obtain
\be
B_{\boldsymbol{k}}=A_{\boldsymbol{k}}e^{2i\phi_{\boldsymbol k}}= \frac{1}{2}\hbar\omega_m\sin^2 \theta_{\boldsymbol k}e^{2i\phi_{\boldsymbol k}},\label{Bk}
\ee
with $\omega_m=\gamma\mu_0M_s$. Here, $M_s$ represents the magnetization of a single sublattice.

By projecting Eq.~(\ref{Hd}) to the magnon particle space $(\psi^+_{\bs k}, \psi^-_{\bs k})$, we obtain
\be
H_{\bs k,-\bs k}^d=\left(\begin{array}{cc}
\Delta_{\boldsymbol{k}}^{++} & i\Delta_{\boldsymbol{k}}^{+-}\\
-i \Delta_{\boldsymbol{k}}^{+-} &-\Delta_{\boldsymbol{k}}^{--}
\end{array}\right),\label{H0d}
\ee
which shows a coupling between the two linear polarized magnon modes. The coupling parameters are
\ber
\Delta_{\boldsymbol{k}}^{++}&=&\Re B_{\boldsymbol{k}}(\gamma_{\boldsymbol{k}}u_{\boldsymbol{k}}^{+}u_{\boldsymbol{k}}^{+}-2v_{\boldsymbol{k}}^{+}u_{\boldsymbol{k}}^{+}+\gamma_{\boldsymbol{k}}v_{\boldsymbol{k}}^{+}v_{\boldsymbol{k}}^{+}),\label{socp}\\
\Delta_{\boldsymbol{k}}^{--}&=&\Re B_{\boldsymbol{k}}(\gamma_{\boldsymbol{k}}u_{\boldsymbol{k}}^{-}u_{\boldsymbol{k}}^{-}+2v_{\boldsymbol{k}}^{-}u_{\boldsymbol{k}}^{-}+\gamma_{\boldsymbol{k}}v_{\boldsymbol{k}}^{-}v_{\boldsymbol{k}}^{-}),\\
\Delta_{\boldsymbol{k}}^{+-}&=&\Im B_{\boldsymbol{k}}(\gamma_{\boldsymbol{k}}u_{\boldsymbol{k}}^{+}u_{\boldsymbol{k}}^{-}+u_{\boldsymbol{k}}^{+}v_{\boldsymbol{k}}^{-}-v_{\boldsymbol{k}}^{+}u_{\boldsymbol{k}}^{-}-\gamma_{\boldsymbol{k}}v_{\boldsymbol{k}}^{+}v_{\boldsymbol{k}}^{-}).\label{soc}
\eer
The total effective non-interacting Hamiltonian $H_{\bs k}$ under the basis of $(\psi^+_{\bs k}, \psi^-_{\bs k})$ thus becomes
\be
H_{\bs k}=\left(\begin{array}{cc}
  \bar\varepsilon_{\bs k}+\delta\varepsilon_{\bs k} & i\Delta_{\boldsymbol{k}}^{+-}\\
  -i\Delta_{\boldsymbol{k}}^{+-} &\bar\varepsilon_{\bs k}-\delta\varepsilon_{\boldsymbol{k}}
\end{array}\right),\label{H0t}
\ee
with $\bar\varepsilon_{\bs k}=(\hbar\omega_{\bs k}^++\hbar\omega_{\bs k}^-+\Delta_{\bs k}^{++}-\Delta_{\bs k}^{--})/2$ and $\delta\varepsilon_{\bs k}=(\hbar\omega_{\bs k}^+-\hbar\omega_{\bs k}^-+\Delta_{\bs k}^{++}+\Delta_{\bs k}^{--})/2$. One see that the magnetic anisotropy supplies a contribution to the spin-orbit field via band splitting $|\hbar\omega_{\bs k}^+-\hbar\omega_{\bs k}^-|$ in $\delta{\varepsilon_{\bs k}}$. Since this band splitting is in subterahertz region, much stronger than the dipolar interaction $|\bs B_{\bs k}|$ in the order of gigahertz, the magnon SOC is dominated by the magnetic anistropy, except around the compensation magnetic field where $|\hbar\omega_{\bs k}^+-\hbar\omega_{\bs k}^-|\simeq 0$.

\subsection{Spin polarized representation}\label{Rspin}
As discussed above, the two magnon eigenstates given by Eq.~(\ref{psipm}) are both linearly polarized, meaning that they do not carry net spin. A unitary transformation into circularly polarized basis can be achieved by
\be
\left(\begin{array}{c}
\alpha_{\boldsymbol{k}}\\
\beta_{\boldsymbol{k}}
\end{array}\right)
={\bf A}\left(\begin{array}{c}
\psi_{\boldsymbol{k}}^{+}\\
\psi_{\boldsymbol{k}}^{-}
\end{array}\right),
\ee
where the transformation matrix is defined as
\be
   {\bf A}=\left(\begin{array}{cc}
\cos\chi & \sin\chi\\
-\sin\chi & \cos\chi
\end{array}\right),\label{transform}
\ee
One can verify that with the parameter $\chi$ given by
\be
\sin2\chi=(u_{\boldsymbol{k}}^{+}u_{\boldsymbol{k}}^{-}+v_{\boldsymbol{k}}^{+}v_{\boldsymbol{k}}^{-})^{-1},
\ee
the two modes $\alpha$ and $\beta$ have one unit spin, but with opposite sign.


In particular, at the compensation magnetic field, we have $\omega_{\bs k}^+=\omega_{\bs k}^-=\omega_{\bs k}$ and $\omega_\delta\ll\omega_{\rm ex}$, which lead to
\ber
u_{\bs k}^+&\simeq& u_{\bs k}^-\simeq u_{\bs k},\\
v_{\bs k}^+&\simeq& -v_{\bs k}^-\simeq v_{\bs k},
\eer
and therefore $\sin2\chi\simeq (u_{\bs k}^2-v_{\bs k}^2)^{-1}=1$.
Under this condition, the transform matrix reduces to
\be
{\bf A}\simeq\left(\begin{array}{cc}
\frac{1}{\sqrt{2}} & \frac{1}{\sqrt{2}}\\
-\frac{1}{\sqrt{2}} & \frac{1}{\sqrt{2}}
\end{array}\right),
\ee
and the SOC coefficients become
\ber
\Delta_{\boldsymbol{k}}^{\pm\pm}&\simeq&\Re B_{\boldsymbol{k}}(\gamma_{\boldsymbol{k}}u^2_{\boldsymbol{k}}-2v_{\boldsymbol{k}}u_{\boldsymbol{k}}+\gamma_{\boldsymbol{k}}v^2_{\boldsymbol{k}}),\\
\Delta_{\boldsymbol{k}}^{+-}&\simeq&\Im B_{\boldsymbol{k}}(\gamma_{\boldsymbol{k}}u^2_{\boldsymbol{k}}-2v_{\boldsymbol{k}}u_{\boldsymbol{k}}+\gamma_{\boldsymbol{k}}v^2_{\boldsymbol{k}}).
\eer
In this spin polarized representation $(\alpha,\beta)$, the effective Hamiltonian reads
\ber
\tilde H_{\bs k}&=&\left(\begin{array}{cc}
\bar\varepsilon_{\bs k} & -\delta\varepsilon_{\boldsymbol{k}}+i\Delta_{\boldsymbol{k}}^{+-}\\
-\delta\varepsilon_{\boldsymbol{k}}-i\Delta_{\boldsymbol{k}}^{+-} & \bar\varepsilon_{\bs k}
\end{array}\right),\nonumber\\
&=&\bar\varepsilon_{\bs k}+\bs h_{\bs k}\cdot\bs \sigma,
\label{tildeH}
\eer
with the effective spin-orbit field $\bs h_{\bs k}=(-\delta \varepsilon_{\bs k},-\Delta_{\bs k}^{+-},0)$.

Up to the first-order of $\omega_{\rm an}/\omega_{\bs k}$, we obtain
\ber
\tilde H_{\bs k}&=&\left(\begin{array}{cc}
\hbar\omega_{\bs k} & -\eta_{\bs k} B_{\bs k}^\ast \\
-\eta_{\bs k} B_{\bs k} & \hbar\omega_{\bs k}
\end{array}\right),\label{approx_H}
\eer
with $\eta_{\bs k}=(\gamma_{\boldsymbol{k}}u^2_{\boldsymbol{k}}-2v_{\boldsymbol{k}}u_{\boldsymbol{k}}+\gamma_{\boldsymbol{k}}v^2_{\boldsymbol{k}})\approx 2 \gamma_{\bs k}\omega_{\rm an}/\omega_{\bs k}$. Eq.~(\ref{approx_H}) is in the same form as the easy-axis case~\cite{Shen20}, because, as aforementioned in Introduction, the magnetic anisotropy and magnetic field together define the effective hard plane ($x$-$y$ plane) and the easy axis ($z$ axis) normal to the plane.

\section{Magnon spin transport}
The spin dynamics of magnons can be described by the semiclassical kinetic equation
\be
\partial_{t}\rho_{\boldsymbol{k}}+i[H_{\boldsymbol{k}},\rho_{\boldsymbol{k}}]
+\frac{1}{2}\{\nabla_{\bs k} H_{\bs k},\nabla \rho_{\bs k}\}=I_{\boldsymbol{k}},\label{kineticE}
\ee
where $\rho_{\bs k}$ is defined as a $2\times 2$ magnon density matrix and the collision integral $I_{\bs k}$ should include all relevant, not only elastic but also inelastic, scattering processes~\cite{Akhiezer68,SSZhang12,Cornelissen16,Liu19,Streib19,Shen2019c,Kamra20,Troncoso20}. By taking into account the large splitting between the two spin bands, the density matrix and Eq.~(\ref{kineticE}) should be written in the representation of $(\psi_{\bs k}^+,\psi_{\bs k}^-)$, especially for an accurate computation of the collision integrals.
The second and third terms on the left side of Eq.~(\ref{kineticE}) describe separately the coherent (quasi)spin precession due to band splitting and the diffusion owing to spatially inhomogenous distribution~\cite{Tokatly16,Kamra20}.

It is important to recall that the density matrix $\rho_{\bs k}(t)$ defined under $(\psi_{\bs k}^+,\psi_{\bs k}^-)$ however does not tell spin information directly. In order to extract the spin polarization, one has to project $\rho_{\bs k}(t)$ into the spin polarized representation via
\be
\tilde\rho_{\bs k}(t)={\bf A}\rho_{\bs k}(t){\bf A}^{\rm T}.
\ee
The magnon spin density then can be read out easily from
\be
s_{\bs k(k)}^i=(1/2)\rm Tr[\tilde\rho_{\bs k(k)}\sigma^i].
\ee

Around the compensation field, the two magnon branches are nearly degenerate, in that case, it is more convenient to write and solve the kinetic equation directly in $(\alpha_{\bs k},\beta_{\bs k})$ representation~\cite{Shen20}
\be
\partial_{t}\tilde\rho_{\boldsymbol{k}}+i[\tilde H_{\boldsymbol{k}},\tilde\rho_{\boldsymbol{k}}]
+\frac{1}{2}\{\nabla_{\bs k} \tilde H_{\bs k},\nabla \tilde\rho_{\bs k}\}=\tilde I_{\boldsymbol{k}}.\label{kineticEb}
\ee
Strictly speaking, different scattering processes will contribute to the dynamics in different ways, relying on their characteristics about the conservation of particle number, spin polarization, momentum and so on~\cite{Shen2019c,Troncoso20}. As a simplified treatment, one may apply the relaxation-time approximation as
\be
\tilde I_{\bs k}=-\frac{1}{\tau}(\tilde\rho_{\bs k}-\tilde\rho_{k}^0),
\ee
where $\tilde \rho_{k}^0$ and $\tau$ represent the quasi-equilibrium density matrix and relaxation time for a specific scattering mechanism~\cite{SSZhang12,Flebus16,Cornelissen16,Shen2019c,Troncoso20}. 

\subsection{Magnon spin relaxation}
After some calculations based on perturbation expansion technique~\cite{Shen:prb2014}, we obtain a drift-diffusion equation~\cite{Shen20,Kamra20}
\be
\partial_t S^i=D \nabla^2 S^i+\epsilon_{ijk}\langle h^j_{\bs k}\rangle S^k-\frac{1}{\tau_s^i}S^i,\label{DDe}
\ee
in which $S^i=\sum_{\bs k}s^i_{\bs k}$ stands for the local spin density and $D=\tau \langle v_{\bs k}^2 \rangle/3$ is the diffusion constant. The first term on the right hand side corresponds to the spin diffusion due to spatial inhomogeneity of the magnon spin density and the second term describes the spin precession around the net effective spin-orbit field $\langle \bs h_{\bs k}\rangle$. Here, $\langle.\rangle$ represents average over all thermally occupied magnon states weighted by the Bose distribution. The magnetic-field dependence of $\langle \bs h_{\bs k}\rangle$ results in a Hanle-type feature, which has been explicitly discussed in Refs.~\cite{Wimmer20,Kamra20}.

\begin{figure}[tp]
  \includegraphics[width=7cm]{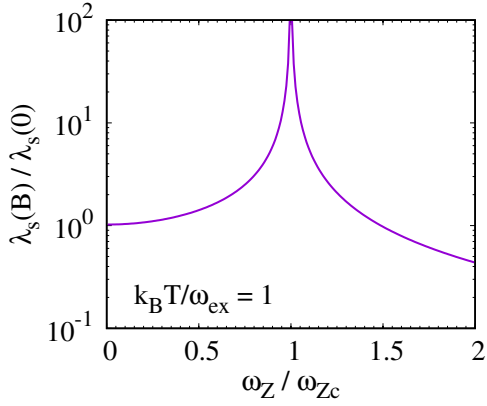}
  \caption{Magnon spin diffusion length as function of external field at temperature $k_BT=\omega_Z$.}
  \label{ls}
\end{figure}

The last term in Eq.~(\ref{DDe}) is the spin relaxation term, which can be caused by various spin non-conserving scattering processes~\cite{Shen20}. Due to the presence of spin-orbit field, the spin-conserving scatterings can also contribute to the spin relaxation via the DP-type mechanism~\cite{DPsrt71}. The spin relaxation time from this machanism reads
\be
(\tau_{s,\rm DP}^{i})^{-1}=\sum_{j\ne i}\tau[\langle(h_{\bs k}^{j})^2\rangle-\langle h_{\bs k}^{j}\rangle^2].\label{DPt}
\ee
In easy-axis AFIs, the magnon spin-orbit field $\bs h_{\bs k}$ is solely from dipole-dipole interaction~\cite{Shen20}. In the present case, the magnetic anisotropy provides an additional contribution. Although this SOC piece is collinear (with only $h_{\bs k}^x$ component), its magnitude varies with frequencies, resulting in a difference between $\langle(h_{\bs k}^{j})^2\rangle$ and $\langle h_{\bs k}^{j}\rangle^2$. Accordingly, the relaxation time $\tau$ should involve the inelastic scatterings, such as magnon-magnon and magnon-phonon scatterings. As the SOC field due to magnetic anisotropy relies on the strength of external field, the spin relaxation rate given by Eq.~(\ref{DPt}) also varies with magnetic field and achieves a minimum at the compensation field. Assuming the diffusion constant $D$ is insensitive to the magnetic field, the magnon spin diffusion length $\lambda_s=\sqrt{D\tau_s}$ will then also vary sharply around the compensation point, as qualitatively shown in Fig.~\ref{ls}.

\subsection{Magnon (inverse) spin Hall effect}
To examine the magnon (inverse) spin Hall effect in the presence of band splitting due to magnetic anisotropy, we next calculate the Berry curvature for spin Hall effect~\cite{sinova04}
\ber
\Omega^{z,\pm}_{x,y}(\bs k)&=&-\frac{2\Im\langle\psi_{\boldsymbol{k}}^{\pm}|\hat v_{x}|\psi_{\boldsymbol{k}}^{\mp}\rangle\langle\psi_{\boldsymbol{k}}^{\mp}|\hat v_{y}^{z}|\psi_{\boldsymbol{k}}^{\pm}\rangle}{(\varepsilon^{\mp}_{\bs k}-\varepsilon^{\pm}_{\bs k})^{2}}, \label{berry}
\eer
where $\varepsilon^{\pm}_{\bs k}$ and $|\psi_{\bf k}^\pm\rangle$ are eigenenergies and wave functions.

For a general Hamiltonian in form of
\be
H_{\bs k}=\varepsilon_{\bs k} +\Delta^x_{\bs k}\sigma_x+\Delta^y_{\bs k} \sigma_y
\ee
one has
\ber
\varepsilon_{\bs k}^\pm&=&\varepsilon_{\bs k}\pm\sqrt{(\Delta^x_{\bs k})^2+(\Delta^y_{\bs k})^2} ,\\
|\psi_{\bf k}^\pm\rangle&=&\frac{1}{\sqrt{2}}\left(\begin{array}{c}
1\\
\pm e^{i\varphi_{\Delta_{\boldsymbol{k}}}}
\end{array}\right).\label{psi}
\eer
The (spin) velocity operators 
\ber
\hat v_{x}&=&v_x^0 +v_x^x\sigma_x+v_x^y\sigma_y,\\
\hat v^z_{y}&=&v_y^0\sigma_{z}.
\eer
where ${v}_{x/y}^0=\partial_{k_{x/y}}\varepsilon_{\bs k}$ and $v_x^{x/y}=\partial_{k_x}\Delta^{x/y}_{\bs k}$. The matrix elements in Eq.~(\ref{berry}) then can be calculated
\ber
\langle\psi_{\boldsymbol{k}}^{\pm}|{\hat v_{x}}|\psi_{\boldsymbol{k}}^{\mp}\rangle
&=&\mp i(v_{x}^{x}\sin\varphi_{\Delta_{\boldsymbol{k}}}-v_{x}^{y}\cos\varphi_{\Delta_{\boldsymbol{k}}}),\\
\langle\psi_{\boldsymbol{k}}^{\mp}|\hat v_{y}^{z}|\psi_{\boldsymbol{k}}^{\pm}\rangle&=&v_y^0.
\eer
By substituting these matrix elements into Eq.~(\ref{berry}), we obtain
\be
\Omega^{z,\pm}_{x,y}(\bs k)=\pm\frac{2v_{y}^0(v_{x}^{x}\sin\varphi_{\Delta_{\boldsymbol{k}}}-v_{x}^{y}\cos\varphi_{\Delta_{\boldsymbol{k}}})}{(\Delta^x_{\bs k})^2+(\Delta^y_{\bs k})^2}.
\ee

Specifically, for the present case, we have
\ber
\varepsilon_{\bs k}&=&(\hbar\omega_{\bs k}^++\hbar\omega_{\bs k}^-+\Delta_{\bs k}^{++}-\Delta_{\bs k}^{--})/2,\\
\Delta_{\bs k}^x&=&-\delta\varepsilon_{\bs k}=-(\hbar\omega_{\bs k}^+-\hbar\omega_{\bs k}^-+\Delta_{\bs k}^{++}+\Delta_{\bs k}^{--})/2,\\
\Delta_{\bs k}^y&=&-\Delta_{\bs k}^{+-}.
\eer
Around the compensation field, the approximate expression of the low energy dispersion relation $\varepsilon_{\bs k}\approx \sqrt{\varepsilon_0^2+c_s^2 k^2}$ gives $v_y^0=({c_{s}^{2}k}/{\varepsilon_{\bs k}})\sin\theta_{\bs k}\sin\phi_{\boldsymbol{k}}$, where $c_s=\omega_{\rm ex}a/2$. And, according to Eqs.~(\ref{tildeH}) and (\ref{approx_H}), we have
\ber
\Delta_{\bs k}^x&\simeq&-\xi_{\bs k}-\zeta_{\bs k}\sin^2 \theta_{\boldsymbol k}\cos(2\phi_{\bs k}),\\
\Delta_{\bs k}^y&\simeq&-\zeta_{\bs k}\sin^2 \theta_{\boldsymbol k}\sin(2\phi_{\bs k}).
\eer
where $\xi_{\bs k}=\hbar(\omega_{\bs k}^+-\omega_{\bs k}^-)/2$ and $\zeta_{\bs k}={\gamma_{\bs k}\hbar\omega_m\omega_{\rm an}}/{\omega_{\bs k}}$ are SOC due to magnetic anisotropy and dipole-dipole interactions, respectively.

Notice that
\be
\tan\varphi_{\Delta_{\boldsymbol{k}}}=\frac{\Delta_{\boldsymbol{k}}^{y}}{\Delta_{\boldsymbol{k}}^{x}}=\frac{2\zeta_{\boldsymbol{k}}k_{x}k_{y}}{k^{2}\xi_{\boldsymbol{k}}+\zeta_{\boldsymbol{k}}(k_{x}^{2}-k_{y}^{2})}.\label{tanphi}
\ee
Very close to the compensation point, the SOC is dominant by the dipolar interaction, i.e., $\zeta_{\bs k}\gg \xi_{\bs k}$. Thus, from Eq.~(\ref{tanphi}), we have $\varphi_{\Delta_{\bs k}}=2\phi_{\bs k}$. The berry curvature then reads
\be
\Omega_{x,y}^{z,\pm}(\boldsymbol{k})=\mp\frac{c_s^2}{\varepsilon_{\bs k}\zeta_{\bs k}}(1+k\frac{\xi_{\boldsymbol{k}}^{\prime}}{\zeta_{\boldsymbol{k}}}\cos^{2}\phi_{\boldsymbol{k}})\frac{\sin^{2}\phi_{\boldsymbol{k}}}{\sin^2\theta_{\bs k}},
\ee
which is globally negative and positive for the upper and lower magnon bands, respectively. This indicates the occurrence of spin Hall effect.

In the opposite limit, the magnetic anisotropy dominates the SOC, i.e., $\xi_{\bs k}\gg \zeta_{\bs k}$, we have
\be
\tan\varphi_{\Delta_{\boldsymbol{k}}}=\frac{\Delta_{\boldsymbol{k}}^{y}}{\Delta_{\boldsymbol{k}}^{x}}\simeq \frac{\zeta_{\boldsymbol{k}}}{\xi_{\boldsymbol{k}}}\sin^{2}\theta_{\boldsymbol{k}}\sin2\phi_{\boldsymbol{k}}\ll1,
\ee
and therefore
\ber
\sin\varphi_{\Delta_{\boldsymbol{k}}}&\simeq& ({\zeta_{\boldsymbol{k}}}/{\xi_{\boldsymbol{k}}})\sin^{2}\theta_{\boldsymbol{k}}\sin2\phi_{\boldsymbol{k}},\\
\cos\varphi_{\Delta_{\boldsymbol{k}}}&\simeq& 1,
\eer
which lead to
\ber
\Omega_{x,y}^{z,\pm}(\boldsymbol{k})&\simeq&\pm\frac{c_s^2}{\varepsilon_{\bs k}\xi_{\bs k}}\left[\frac{\zeta_{\boldsymbol{k}}}{\xi_{\bs k}}+(\frac{k\zeta_{\boldsymbol{k}}^{\prime}}{\xi_{\bs k}}-2\frac{\zeta_{\boldsymbol{k}}}{\xi_{\bs k}})\sin^{2}\theta_{\boldsymbol{k}}\cos^{2}\phi_{\boldsymbol{k}}\right]\nonumber\\
      &&\times \sin^{2}\theta_{\boldsymbol{k}}\sin^{2}\phi_{\boldsymbol{k}}.
\eer
One sees that the Berry curvature reduces with ${\zeta_{\boldsymbol{k}}}/{\xi_{\bs k}}$, meaning the suppression of the spin Hall effect by the magnetic anisotroy. This is because of the collinear nature of the SOC due to magnetic anisotropy.

\section{Influence of DMI and in-plane anisotropy}
In some easy-plane antiferromagnetic magnets like $\alpha$-Fe$_2$O$_3$, there is a zero-field magnetization induced by DMI. To examine the consequence of DMI, we describe it by
\be
H^{\rm DM}=D\sum_{{\langle i,j\rangle}^\prime}\hat{x}\cdot\boldsymbol{S}_{ai}\times\boldsymbol{S}_{dj},
\ee
where only those DMI-active bonds are counted in the summation. The DMI then leads to an additional energy
\be
E^{\rm DM}=-Nz'DS^{2}\sin2\theta. \label{EDMI}
\ee
where $z'$ stands for the number of neighboring ions connected by DMI. The condition of the equilibrium canting angle then can be derived by including Eq.(\ref{EDMI}) into Eq.~(\ref{totalE}) as
\be
\omega_{Z}\cos\theta+\omega_{{\rm DM}}\cos2\theta=\omega_{{\rm ex}}\sin2\theta,
\ee
with $\omega_{\rm DM}=z'DS/\hbar$. After some calculations following the techniques introduced in Sec.~\ref{smodel}, we find its contribution to magnon Hamiltonian can be included by the substitutions
\ber
    {\cal A}	&\to&	{\cal A}+\hbar\omega_{{\rm DM}}\tan\theta,\\
    {\cal C}_{\boldsymbol{k}}	&\to&\hbar\gamma_{\boldsymbol{k}}\omega_{{\rm ex}}\sin^{2}\theta\left(1-\frac{\gamma_{\boldsymbol{k}}^{\prime}\omega_{{\rm DM}}}{\gamma_{\boldsymbol{k}}\omega_{{\rm ex}}}\cot\theta\right).
\eer
In reality, only part of the exchange interacting bonds are involved in the DMI, which means in general $\gamma_{\bs k}\ne\gamma_{\bs k}^\prime$. This makes ${\cal C}_{\boldsymbol{k}}/{\gamma_{\bs k}}$ no longer a constant. As a result, $\omega_{\bs k}^+=\omega_{\bs k}^-$ is not able to satisfy in the entire Brillouin zone for any magnetic field. Namely, no compensation field is allowed and the DMI provides an effective spin-orbit field at any external magnetic field, from which it can affect the magnon spin relaxation and spin Hall effect.

In biaxial antiferromagnets like the intensively studied material, NiO~\cite{WangHL2014,Lin16,YiWang2019}, there is an easy axis within the easy plane. This effect can be included by an in-plane magnetic anisotropy term~\cite{NiO72,Rezende19}
\be
H^{\rm in}=\sum_i K'(S_{ai}^{z})^{2}+K'(S_{di}^{z})^{2},
\ee
with anisotropy parameter $K'<0$. For simplicity, we here assume the in-plane easy axis is along $z$-direction, i.e., perpendicular to the applied field. This term gives an enengy
\be
E^{\rm in}=2NK'S^{2}\cos^{2}\theta.
\ee
By taking this term into account, we find the condition of the equilibrium (without DMI) canting angle
\be
\sin\theta=\frac{\omega_{Z}}{2(\omega_{{\rm ex}}-\omega_{{\rm an}}^{\prime})},
\ee
with $\omega_{{\rm an}}^{\prime}=|K'|S/\hbar$ and the corrections to the magnon Hamiltonian can be included in Hamiltonian~(\ref{H00}) via the replacement
\ber
{\cal A}&\to&{\cal A}+\hbar\omega_{{\rm an}}^{\prime}(2-3\sin^{2}\theta),\\
{\cal A}'&\to&{\cal A}'+\hbar\omega_{{\rm an}}^{\prime}\sin^{2}\theta.
\eer
 Since the corrections are moment independent, the compensation features will survive. 

\section{Summary}
In summary, we study the magnon spin transport in easy-plane antiferromagnetic insulators under an in-plane magnetic field. From the analysis on the influence of the magnetic field, we find the two magnon branches becomes degenerate at a compensation magnetic field, making the easy-plane antiferromagnet equivalent to the uniaxial easy-axis antiferromagnets. At this compensation condition, magnon spin-orbit coupling due to dipolar interaction results in magnon (inverse) spin Hall effect and D'yakonove Perel'-type spin relaxation. The compensation feature is found to survive in biaxial easy-plane systems but will be removed by Dzyaloshinskii–Moriya interaction. Far away from the compensation magnetic field, the magnon spin-orbit coupling is dominated by the magnetic anisotropy, where the magnon (inverse) spin Hall effect is suppressed. These results are expected to be applicable in synthetic antiferromagnets, in which the larger magnetic moments of the artificial spin elements benefit the enhancement of the predicted dipolar-induced spin-orbit effects.

\begin{acknowledgments}
  This work is supported by the National Natural Science Foundation of China (Grants No.11974047).
\end{acknowledgments}


{\bf{DATA AVAILABILITY}\\}\\
The data that support the findings of this study are available from the corresponding author upon reasonable request.


\end{document}